\documentclass[pre,superscriptaddress,amsmath,amssymb,nofootinbib,floatfix,onecolumn,notitlepage]{revtex4-1}
\usepackage{amsfonts, amsmath, amssymb, amsthm, array, eucal, mathtools, xfrac,
booktabs, array, color, natbib, graphicx, siunitx, tikz, pgfplots,subcaption}
\pgfplotsset{compat=1.8, every axis/.append style={font=\footnotesize}}
\usetikzlibrary{shapes,patterns,arrows,fadings,shadows}
\definecolor{midori}{rgb}{0.1,0.6,0}
\pgfplotsset{
    colormap={parula}{
        rgb=(0.2081,0.1663,0.5292)
        rgb=(0.2116,0.1898,0.5777)
        rgb=(0.2123,0.2138,0.627)
        rgb=(0.2081,0.2386,0.6771)
        rgb=(0.1959,0.2645,0.7279)
        rgb=(0.1707,0.2919,0.7792)
        rgb=(0.1253,0.3242,0.8303)
        rgb=(0.0591,0.3598,0.8683)
        rgb=(0.0117,0.3875,0.882)
        rgb=(0.006,0.4086,0.8828)
        rgb=(0.0165,0.4266,0.8786)
        rgb=(0.0329,0.443,0.872)
        rgb=(0.0498,0.4586,0.8641)
        rgb=(0.0629,0.4737,0.8554)
        rgb=(0.0723,0.4887,0.8467)
        rgb=(0.0779,0.504,0.8384)
        rgb=(0.0793,0.52,0.8312)
        rgb=(0.0749,0.5375,0.8263)
        rgb=(0.0641,0.557,0.824)
        rgb=(0.0488,0.5772,0.8228)
        rgb=(0.0343,0.5966,0.8199)
        rgb=(0.0265,0.6137,0.8135)
        rgb=(0.0239,0.6287,0.8038)
        rgb=(0.0231,0.6418,0.7913)
        rgb=(0.0228,0.6535,0.7768)
        rgb=(0.0267,0.6642,0.7607)
        rgb=(0.0384,0.6743,0.7436)
        rgb=(0.059,0.6838,0.7254)
        rgb=(0.0843,0.6928,0.7062)
        rgb=(0.1133,0.7015,0.6859)
        rgb=(0.1453,0.7098,0.6646)
        rgb=(0.1801,0.7177,0.6424)
        rgb=(0.2178,0.725,0.6193)
        rgb=(0.2586,0.7317,0.5954)
        rgb=(0.3022,0.7376,0.5712)
        rgb=(0.3482,0.7424,0.5473)
        rgb=(0.3953,0.7459,0.5244)
        rgb=(0.442,0.7481,0.5033)
        rgb=(0.4871,0.7491,0.484)
        rgb=(0.53,0.7491,0.4661)
        rgb=(0.5709,0.7485,0.4494)
        rgb=(0.6099,0.7473,0.4337)
        rgb=(0.6473,0.7456,0.4188)
        rgb=(0.6834,0.7435,0.4044)
        rgb=(0.7184,0.7411,0.3905)
        rgb=(0.7525,0.7384,0.3768)
        rgb=(0.7858,0.7356,0.3633)
        rgb=(0.8185,0.7327,0.3498)
        rgb=(0.8507,0.7299,0.336)
        rgb=(0.8824,0.7274,0.3217)
        rgb=(0.9139,0.7258,0.3063)
        rgb=(0.945,0.7261,0.2886)
        rgb=(0.9739,0.7314,0.2666)
        rgb=(0.9938,0.7455,0.2403)
        rgb=(0.999,0.7653,0.2164)
        rgb=(0.9955,0.7861,0.1967)
        rgb=(0.988,0.8066,0.1794)
        rgb=(0.9789,0.8271,0.1633)
        rgb=(0.9697,0.8481,0.1475)
        rgb=(0.9626,0.8705,0.1309)
        rgb=(0.9589,0.8949,0.1132)
        rgb=(0.9598,0.9218,0.0948)
        rgb=(0.9661,0.9514,0.0755)
    }
}
\newlength\fwidth
\newlength\fheight

\begin{document}

\title{Fake $\mu$s: A cautionary tail of shear-thinning locomotion}
\author{Thomas D. Montenegro-Johnson}
\affiliation{School of Mathematics, University of Birmingham, Edgbaston, Birmingham, B15 2TT}
\date{\today}
\begin{abstract}
{ Propulsion through fluids is a key component in the life cycle of many microbes, be it in development, infection, or simply finding nutrients. In systems of biomedical relevence, this propulsion is often through polymer suspensions that endow the fluid with complex non-Newtonian properties, such as shear-thinning and viscoelastic behaviour. Due to the complexity of 3D non-Newtonian modelling,} 2D undulatory propulsion {has recently been} extensively-studied as a means of garnering physical intuition for {these systems. However, while} streamlines, swimming speeds, and swimmer trajectories are strikingly similar in {2D and 3D Newtonian calculations, behaviour in non-Newtonian} fluids depends upon flow derivatives, such as the shear rate, which are radically different. { Taking shear-thinning as an example rheology, prevalent in biological fluids such as physiological mucus, this communication demonstrates how failing to account for} this difference can misguide our understanding of 3D non-Newtonian swimming.
\end{abstract}

\maketitle

\section{Introduction}

Microscale undulatory swimming is a key component of many important biological processes, such as mammalian reproduction~\citep{gaffney2011mammalian} {and Lyme disease infection~\citep{charon2012unique}, the most commonly-reported vectorborne illness in the US. Through waving} a slender tail (flagellum), undulatory swimmers such as sperm generate propulsive force~\citep{hancock1953self, gray1955propulsion, lighthill1976flagellar} which results in forward motion~\citep{lauga2009hydrodynamics}. { In both of these examples, as with other biomedically-relevant systems, the microswimmer must progress} through a non-Newtonian fluid, physiological mucus~\citep{hwang1969rheological} {for the sperm, and the mammalian dermis for the spirochete}. 

In 1951, G. I. Taylor initiated the fluid dynamical study of undulatory propulsion by demonstrating that a two-dimensional (2D), infinite sheet could swim by propagating a travelling wave along its body~\citep{taylor1951analysis}. A year later, Taylor extended this 2D analysis to an infinite cylinder undergoing planar undulations~\citep{taylor1952action}. Subsequently, a variety of Newtonian fluid studies were undertaken that led to the development of slender body theory~\citep{hancock1953self, lighthill1976flagellar, johnson1979improved} and its algebraic approximation resistive force theory~\citep{gray1955propulsion, lighthill1976flagellar}, while additional studies incorporated nonlocal interactions between the flagellum and swimmer bodies~\citep{higdon1979hydrodynamic}, boundary features~\citep{fauci1995sperm, smith2009human}, and the elastohydrodynamics of the undulating flagellum~\citep{machin1958wave, hines1978bend, hines1979bend, gadelha2010nonlinear, montenegro2015spermatozoa}.

{ Owing to the biomedical importance of non-Newtonian undulatory propulsion, there has recently been an extensive research program to understand the effects of non-Newtonian rheology upon undulatory propulsion.} Barring a few exceptions~\citep{fu2007theory,fu2009swimming}, these studies have typically been 2D; they include computational studies in shear-thinning fluids with finite elements~\citep{montenegro2012modelling, montenegro2013physics}, finite volume methods~\citep{li2015undulatory}, immersed boundary simulations in viscoelastic fluids~\citep{teran2010viscoelastic, thomases2014mechanisms}, and a significant number of asymptotic studies utilising Taylor's swimming sheet approach to study shear-thinning~\citep{velez2013waving}, viscoelastic~\citep{lauga2007propulsion, elfring2010two, riley2014enhanced, riley2015small}, two-fluid~\citep{mirbagheri2016helicobacter}, yield-stress~\citep{hewitttaylor}, and transversely isotropic fluids~\citep{cupples2017viscous}, as well as liquid crystals~\citep{krieger2015microscale}. The non-Newtonian aspects of these flows typically depend upon flow derivative quantities, such as the shear rate or normal stresses. This dependence motivates the question as to whether these 2D flow derivatives are representative of those in equivalent 3D models, and whether any differences qualitatively affect the non-Newtonian flow.  

This paper will focus upon the effects of differences in flow derivative calculation for microscale undulatory swimming in shear-thinning {rheology, an important non-Newtonian behaviour that has been observed in physiological mucus~\citep{lai2009micro}}. Recent 2D computational studies~\citep{montenegro2012modelling,montenegro2013physics,li2015undulatory} have concluded that gradients in fluid viscosity, creating a corridor of thinned fluid surrounding the swimmer, are key to understanding the effects of shear-thinning on swimmer velocity. Similarly, G{\'o}mez et al.~\cite{gomez2017helical} found that a ``confinement effect'' arising from a corridor of thinned fluid surrounding an artificial bacterial swimmer was responsible for a significantly enhanced velocity in shear-thinning fluids. Since this corridor of thinned fluid arises from local fluid shear, it is thus critical to understand the shear rates of the surrounding flow. However, the form of these shear rates is dramatically different for 2D vs 3D undulatory swimmers, owing to the inability of the fluid to ``flow over'' a 2D sheet.

\subsection{Equations of flow}

For shear-rate dependent fluids at microscopic scales, flow dynamics is well-modelled by the generalised Stokes flow equations, 
\begin{equation}
\boldsymbol{\nabla}\cdot(\mu\boldsymbol{\varepsilon}(\mathbf{u})) =
\boldsymbol{\nabla}p, \ \ \boldsymbol{\nabla}\cdot\mathbf{u} =
0, \ \  \boldsymbol{\varepsilon}(\mathbf{u}) = \boldsymbol{\nabla}\mathbf{u} +
\boldsymbol{\nabla}\mathbf{u}^T\!\!,
\end{equation}
for fluid velocity $\mathbf{u}$, pressure $p$ and viscosity $\mu$. For Newtonian fluids, the viscosity $\mu$ is constant, but for rate-dependent fluids, $\mu$ is a function of the shear rate, $\dot{\gamma} = \sqrt{\boldsymbol{\varepsilon}(\mathbf{u}):\boldsymbol{\varepsilon}(\mathbf{u})/2}$. Thus, innacurate calculation of flow derivatives can lead to incorrect viscosity fields or ``fake $\mu$s'', which can misguide our understanding of non-Newtonian rheological effects.

For shear-thinning fluids, such as physiological mucus, the viscosity dependence upon the shear rate may be modelled by the Carreau law~\citep{carreau1972rheological},
\begin{equation}
\mu(\dot{\gamma}) = \mu_\infty + (\mu_0 - \mu_\infty)(1 +
(\lambda \dot{\gamma})^2)^{(n-1)/2},
\end{equation}
where $\mu_0,\ \mu_\infty$ are the fluid viscosity at zero and infinite shear rates respectively, and the time constant $\lambda$ and power-law index $n$ affect the speed of the monotonic decrease from $\mu_0$ to $\mu_\infty$ as shear rate $\dot{\gamma}$ increases. In dimensionless form, the viscosity is given,
\begin{equation}
\mu(\dot{\gamma}) = \mu_s + (1 - \mu_s)(1 +
(\mathrm{Cu} \dot{\gamma})^2)^{(n-1)/2},
\end{equation}
and we see that the important parameters are $n$, the viscosity ratio $\mu_s = \mu_\infty/\mu_0$ and Carreau number $\mathrm{Cu} = \omega\lambda$, for flagellar waveform angular frequency $\omega$. 

{Thus for shear-thinning, as with most non-Newtonian fluids}, flow dynamics depends nonlinearly upon the flow velocity derivatives, {and as such it is important to choose a model which accurately represents these derivatives. However, in the following argument we will neglect this nonlinear flow feedback and consider Newtonian base cases in 2D and 3D, which can give an approximate indication of the ``zeroth-order'' behaviour~\citep{goddard1976tensile} of the non-Newtonian system. Our justification for this approach in the case of shear-thinning runs as follows,} 
\begin{enumerate}
\item For certain biological systems of interest, the wave amplitude is comparatively small; for instance, the wave amplitude of the nematode worm \textit{C. Elegans} is approximately a tenth of the wavelength~\citep{Shen2011}.
\item The 1\textsuperscript{st}- and 2\textsuperscript{nd}-order solutions for a waving sheet in a shear-thinning fluid are identical to the Newtonian case, irrespective of the extent of shear-thinning rheology, with changes to propulsive velocity only appearing at 4\textsuperscript{th}-order in amplitude~\citep{velez2013waving}.
\item Indeed, for large-amplitude 2D simulations of a waving sheet in shear-thinning fluid, the shear rate is well approximated by the small-amplitude Newtonian solution for a range of viscosity parameters~\citep{li2015undulatory}.
\item {A scaling law for power expenditure in shear-thinning fluids derived by Li and Ardekani~\cite{li2015undulatory} in their 2D work, which relies on the flow shear rate being largely insensitive to the degree of shear-thinning rheology, has subsequently been verified experimentally by Gagnon and Arratia~\cite{gagnon2016cost}.}
\end{enumerate}
{ Thus, differences between shear rates as calculatated via small-amplitude Newtonian models can give good qualitative understanding of realistic systems involving shear-thinning rheology.} Our argument { thus begins with} Taylor's classic results for the flow surrounding a small-amplitude swimming sheet~\citep{taylor1951analysis} and a comparable waving cylinder~\citep{taylor1952action}; In the 2D case, the shear rate takes a maximum away from the swimmer, whereas in the 3D case this maximum is on the swimmer. This difference in the underlying Newtonian case has, to good approximation~\citep{li2015undulatory}, viscosity in a shear-thinning fluid changing from thick to thin to thick moving away from the swimmer in 2D, and simply thin to thick in 3D; the 2D model results in a spurious understanding of the 3D viscosity field, and in particular the pseudo-confinement effect discussed by Li and Ardekani~\cite{li2015undulatory} and demonstrated experimentally by G{\'o}mez et al.~\cite{gomez2017helical}.

\section{Results}
\subsection{Small-amplitude modelling}

\begin{figure*}[t]
    \begin{center}
	\includegraphics{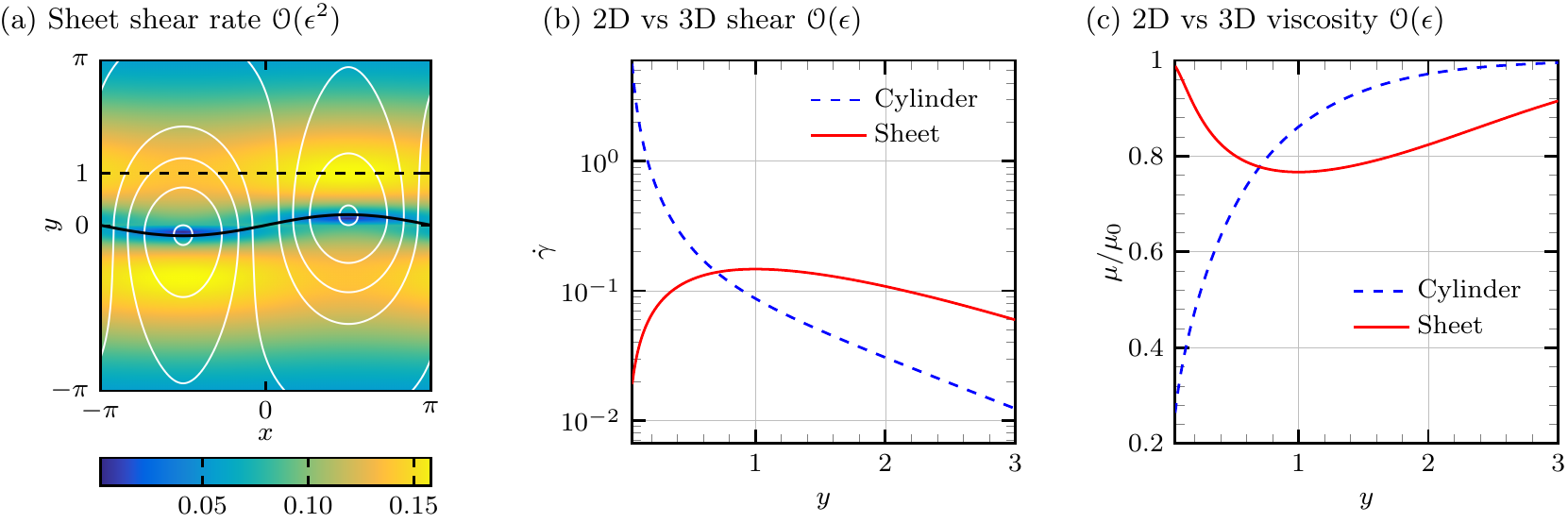}
	\caption{(a) 2\textsuperscript{nd}-order shear rate of flow around a swimming sheet, showing maximum shear ``off'' the swimmer surface. {Flow streamlines are shown in white.} (b) 1\textsuperscript{st}-order shear rates as a function of distance from the swimmer in the plane of beating for the flow around a cylinder with radius $a = 0.05$ and a sheet, with small-amplitude undulations $b = 0.2$. (c) Effective viscosity relative to $\mu_0$ as a function of distance away from the sheet/filament for models approximating a nematode swimming in 300 ppm Xanthum gum solution~\citep{gagnon2016cost}, with $\lambda = 1.2\,\mathrm{s},\ n = 0.7$, and beat frequency $f =2\,\mathrm{Hz}$, so that $\mathrm{Cu} = 4\pi\lambda\approx 15$.}
	\label{fig:1}
	\end{center}
    \vspace{-0.5cm}
\end{figure*}

For a swimming sheet, we prescribe the centreline kinematics $z = s,\ y = b\sin(s - t)$, for $s\in(0,2\pi)$. The flow velocity to first order is then~\citep{taylor1951analysis},
\begin{equation}
    v = b(1 + y)e^{-y}\cos(s-t), \quad w = bye^{-y}\sin(s-t),
\end{equation}
and the 2D shear rate $\dot{\gamma} = [2w_z^2 + (w_y + v_z)^2 + 2v_y^2]^{1/2}$, whereupon substitution of the flow velocity derivatives reveals the somewhat surprising result that the first-order shear rate $\dot{\gamma}$ is independent of position $s$,
\begin{equation}
    \dot{\gamma} = 2b y e^{-y}, \ \ \dot{\gamma}_y = 2b e^{-y}(1 - y) = 0, \
\ \Rightarrow \ \ y_m = 1.
\end{equation}
Thus, at first order in 2D, the shear rate is zero on the swimmer, increases to a maximum and then decays exponentially, with the maximum located at $y_m$ a distance of one wavenumber of the swimmer travelling wave. This result is essentially unaffected at second-order, where for convenience we examine the square of the fluid shear rate,
\begin{equation}
    \dot{\gamma}^2 = 4b^2[y^2e^{-2y} + 2by(2y - 1)e^{-3y}\sin(s - t) 
    + b^2e^{-4y}(4y^2 - 4y + 1)],
\end{equation}
and looking for the maxima with respect to $y$ we still see the maximum in the $y$-direction for fixed $s$ ``off'' the swimmer at $y = 1$,  with global maxima over the wavecrests and minima over wave troughs corresponding to the thickened ``plugs'' of fluid visible in viscosity fields simulated by both Montenegro-Johnson et al.~\cite{montenegro2013physics} and Li and Ardekani~\cite{li2015undulatory}, as well as in recent work in Bingham fluids~\citep{hewitttaylor}. The second order shear rate is plotted (along with streamlines) in figure~\ref{fig:1}a.

For an infinite cylinder of small radius $a$ undergoing the same centreline kinematics, the flow velocity on the plane of the beating is given by~\citep{taylor1952action, fu2009swimming}, 
\begin{subequations}
    \begin{align}
        u &= b\cos\phi[ArK_1(r) + BK_2(r) + CK_0(r)], \\
        v &= b\sin\phi[BK_2(r) - CK_0(r)], \\
        w &= b\cos\phi[ArK_0(r) + (B + C - A)K_1(r)],
    \end{align}
\end{subequations}
for $K_i$ modified Bessel functions of the second kind, with
\begin{gather}
        A = \left(K_0(a) + a K_1(a)\left[\frac{1}{2} +
        \frac{1}{2}\frac{K_0(a)}{K_2{a}} - 
        \frac{K_0(a)^2}{K_1(a)^2}\right]\right)^{-1}\!\!\!\!\!\!\!, 
        \quad B = -\frac{Aa}{2}\frac{K_1(a)}{K_2(a)}, \quad
        C = \frac{1 - AaK_1(a)/2}{K_0(a)}.
\end{gather}
We make the comparison with the 2D shear rate in the plane of beating $\phi = 0$, $r = y$, where by symmetry~\citep{montenegro2016flow} the shear rate is $\dot{\gamma} = (2u_y^2 + w_y^2 + 2(v_\phi/y)^2)^{1/2}$. Since
\begin{subequations}
\begin{align}
    u_y    &= b[AK_1(y) + AyK_1^{\prime}(y) + BK_2^{\prime}(y) +
CK_0^{\prime}(y)],\\
    w_y    &= b[AK_0(y) + AyK_0^{\prime}(y) + (B\! +\! C\! -\!
A)K_1^{\prime}(y)],\\
    v_\phi &= b[BK_2(y) - CK_0(y)],
\end{align}
\end{subequations}
with the derivatives given by,
\begin{equation}
K_0^{\prime}(y) = -K_1(y), \quad
K_1^{\prime}(y) = -K_0(y) - K_1(y)/y, \quad
K_2^{\prime}(y) = -K_1(y) - K_2(y)/y,
\end{equation}
we can calculate $\dot{\gamma}$ for a range of flagellum radii, for which we see that the maximum value of shear rate is clearly on the flagellum itself (fig.~\ref{fig:1}b). 

The stark contrast between the cylinder and the sheet shear rates is shown in figure~\ref{fig:1}b. Not only are the maxima located at different distances from the body, but the maximum shear rate is significantly higher for the cylinder. The simple explanation is that fluid may `flow around' the cylinder but not the sheet, giving a very large out of plane contribution to the shear rate at the boundary in 3D, despite the fact that the 2D and 3D models produce similar flow streamlines in the plane of beating and similar swimmer trajectories. 

Under the assumption that the Newtonian and non-Newtonian shear rates are approximately equal~\citep{li2015undulatory,gagnon2016cost}, we can examine the difference in viscosity fields between a 2D sheet and a 3D cylinder exhibiting the same kinematics. V{\'e}lez-Cordero and Lauga~\cite{velez2013waving} noted that rheological data of shear-thinning fluids typically indicated that $\mu_0 \gg \mu_\infty$, so that to a reasonable approximation,
\begin{equation}
\mu(\dot{\gamma}) = (1 + (\mathrm{Cu}
\dot{\gamma})^2)^{(n-1)/2},
\end{equation}
where $\mu$ is normalised by the zero shear rate viscosity. For nematode worms in 300 ppm Xanthum gum solutions~\citep{gagnon2016cost}, $\lambda = 1.2\,\mathrm{s}$ and $n = 0.7$, while the beat frequency is around $2\,\mathrm{Hz}$ giving a Carreau number $\mathrm{Cu} = 4\pi\lambda\approx 15$. Based upon the shear rates in figure~\ref{fig:1}b, the relative viscosity of such a fluid as a function of the distance from the swimmer is shown in figure~\ref{fig:1}c. The difference between the 2D and 3D swimmers is dramatic; at the surface of the cylinder, the viscosity is approximately 5 times less viscous than the sheet, and the fluid returns to 90\% of the zero shear rate viscosity at a distance of 1.2 for the cylinder compared to 2.8 for the sheet. As Gagnon and Arratia~\cite{gagnon2016cost} note, power expenditure in a shear-thinning fluid is approximately proportional to the fluid viscosity, and as both Li and Ardekani~\cite{li2015undulatory} and G{\'o}mez et al.~\cite{gomez2017helical} note, swimming may be enhanced, for fixed kinematics, via pseudo-wall effects of being in a channel of thinned fluid, so that calculations of power expenditure and swimming enhancement are likely to be inaccurate if 2D modelling is employed.



\subsection{Nematode worm modelling, 2D vs 3D}

Does this effect persist for finite-length swimmers? We generate a smooth-time representation of the planar beating of \textit{C. elegans} by fitting the experimentally-extracted~\citep{gagnon2014undulatory,gagnon2016cost} curvature $\kappa(s,t)$ of the worm's body to the function~\citep{thomases2014mechanisms}
\begin{equation}
    \kappa(s,t) = \left[c_1 - c_2s\right]\left[\cos\left(\omega t +
            c_3\pi(L - s) + c_4\right)\right]\hfill
    \label{eq:curv_fit}
\end{equation}
where $\omega$ is the angular frequency of the swimmer beat, $s$ is the length along the swimmer from the head, $t$ is time and $L$ is the length of the swimmer. Fitting is performed using the Matlab routine \texttt{lsqcurvefit}, which solves nonlinear curve-fitting in a least-squares sense. 

\begin{figure}[t]
    \begin{center}
	\includegraphics{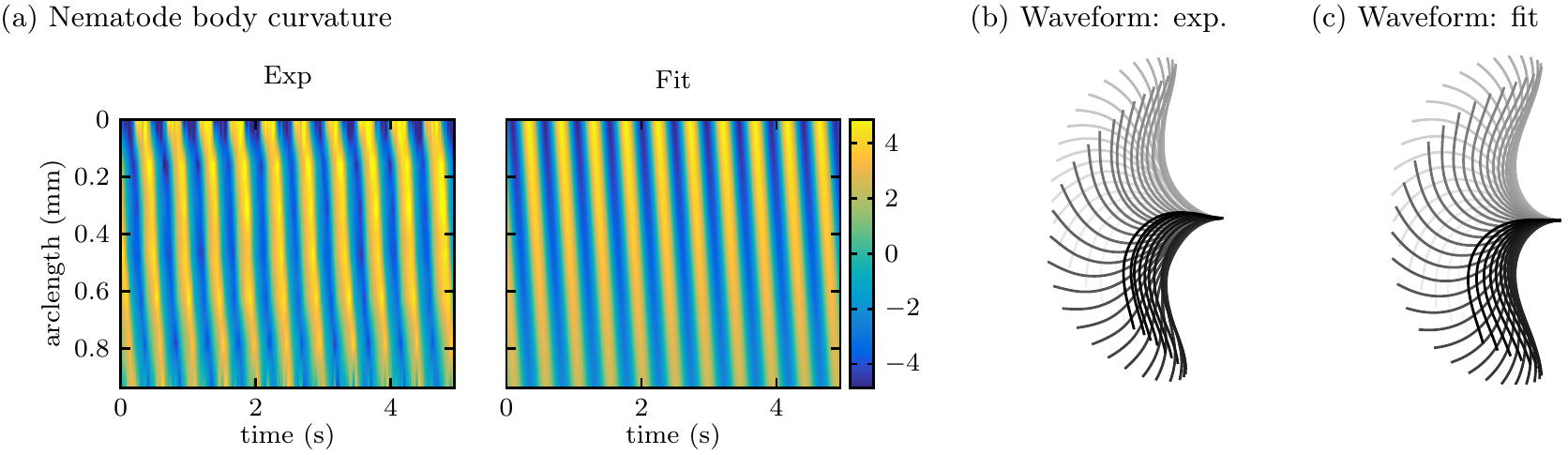}   
        \caption{Numerical fitting of swimmer kinematics. Curvature of an     example swimmer as (a) extracted from microscopy data and (b) fitted to equation~\eqref{eq:curv_fit}. (c) The experimentally extracted waveform in the body frame is shown next to the numerical waveform (d) achieved through integrating the fitted curvature (b) twice.} 
	\label{fig:curvature} 
	\end{center}
    \vspace{-0.5cm}
\end{figure}

The swimmer curvature is shown in comparison to the numerical fit in figure~\ref{fig:curvature}a. This curvature is then integrated with respect to arclength to give the body centreline tangent angle with respect to the head, $\psi(s,t)$, and integrated again to provide the waveform $\mathbf{x}_c(s,t)$ (fig.~\ref{fig:curvature}b,c) in the `body frame', where the head neither rotates or translates,
\begin{equation}
\begin{gathered}
    \psi(s,t) = \int_0^L{\kappa(s,t)}\,\mathrm{d}s, \quad 
    \mathbf{x}_c(s,t) = \int_0^L\mathbf{t(\psi)}\,\mathrm{d}s, \quad
    \mathbf{u}_c(s,t) = \int_0^L\dot{\psi}\mathbf{n(\psi)}\,\mathrm{d}s, \\  
    \mathbf{t} = \left[\cos\psi, \sin\psi, 0\right]^\intercal \quad
    \mathbf{n} = \left[-\sin\psi, \cos\psi, 0\right]^\intercal 
\label{eq:centerline}
\end{gathered}
\end{equation}
where the body frame velocity $\mathbf{u}_c$ is obtained by differentiating the centreline with respect to time.

We model the swimmer as a line distribution of regularised stokeslets in 2D~\citep{cortez2002method} and 3D~\citep{cortez2005method}. For a regularised driving force located at $\mathbf{x}$, $\varphi_\epsilon(r)\mathbf{f}$, and any point in the flow domain $\mathbf{y}$, we have in the two-dimensional regularised stokeslet,
\begin{equation}
    \varphi^{2D}_\epsilon(r) = \frac{3\epsilon^3}{2\pi r_\epsilon^5}, \quad r_i =
y_i-x_i, \quad r_\epsilon^2 =
    r^2 + \epsilon^2, \quad   \mathbf{S}_{ij}^{\scriptsize{2D}} =
\frac{1}{4\pi}\Big(-\delta_{ij}\left[\ln(r_\epsilon
+ \epsilon) - \frac{\epsilon r_\epsilon +
2\epsilon^2}{r_\epsilon^2 + \epsilon r_\epsilon}\right] +
\frac{r_i r_j[r_\epsilon + 2\epsilon]}{r_\epsilon[r_\epsilon +
\epsilon]^2}\Big),
\end{equation}
and the three-dimensional regularised stokeslet,
\begin{equation}
    \varphi^{3D}_\epsilon(r) = \frac{15\epsilon^4}{8\pi r_\epsilon^7}, \quad r_\epsilon^2 =
    r^2 + \epsilon^2, \quad
    \mathbf{S}_{ij}^{\scriptsize{3D}} = \frac{\delta_{ij}(r^2 +
    2\epsilon^2) + r_i r_j}{8\pi r_\epsilon^3}.
\end{equation}
The flow velocity $\mathbf{u}(\mathbf{x}_0)$ at a point $\mathbf{x}_0$ is then given as a line integral of stokeslets $\mathbf{S}(\mathbf{x}_c,\mathbf{x}_0)$ weighted by the unknown force-per-unit length $\mathbf{f}$ that the swimmer exerts upon the fluid. To find this unknown force, we discretise the swimmer into $N$ elements $E_j$ with constant force per unit length~\citep{smith2009boundary}, yielding the matrix system 
\begin{equation} 
\mathbf{u}_c(\mathbf{x}_i) + \mathbf{U} +
\mathbf{x_0}\wedge\boldsymbol{\Omega} =
\sum_{j=1}^N\mathbf{f}_j\int_{E_j}\mathbf{S}(\mathbf{x}_i,\mathbf{x}_j)\,\mathrm{d}s.
\end{equation} 
for $\mathbf{x}_{i,j}$ the element centroids.  The unknown swimming translational and angular velocities $\mathbf{U},\ \boldsymbol{\Omega}$ are found by specifying that the swimmer exerts zero net force and torque on the fluid. 

\begin{figure*}[tb]
    \begin{center}
    \includegraphics{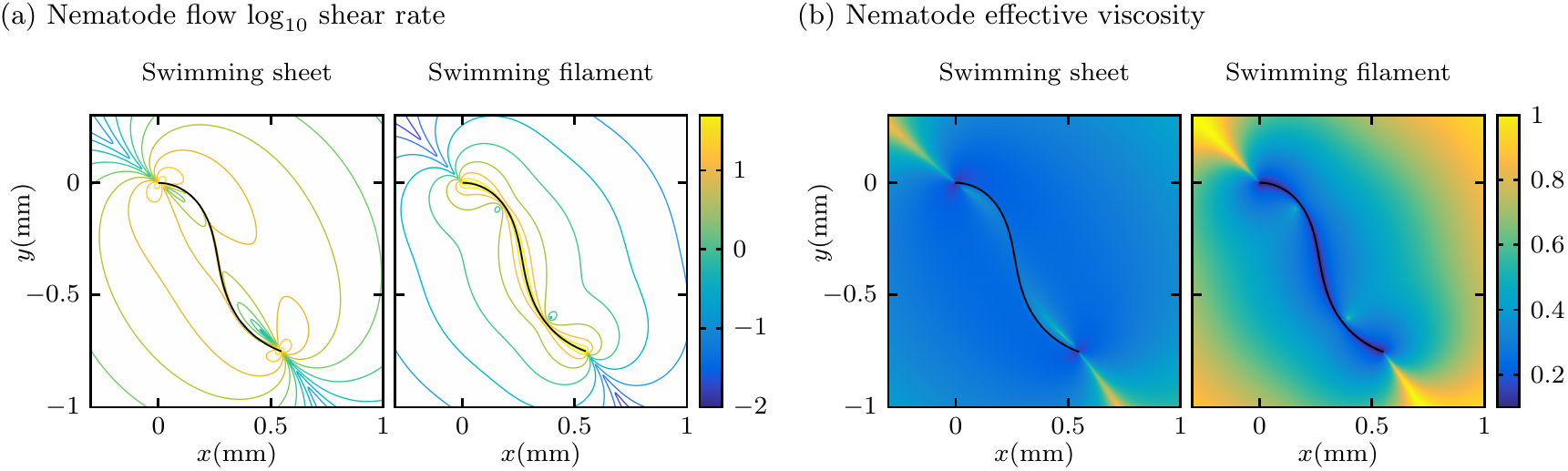}
    \caption{{(a) Logarithm of the shear rate $\log_{10}\dot{\gamma}$} of flow surrounding a 2D sheet and 3D filament where kinematics have been fitted to experimental data of \textit{C. Elegans} centreline swimming. This finite-length solution displays similar characteristics to the first-order infinite solutions fig.~\ref{fig:1}b and (b) the effective viscosity fields associated with these shear rates in 300~ppm Xanthum gum solution~\citep{gagnon2016cost}, showing striking differences.}
    \label{fig:worm_shear}
    \end{center}
    \vspace{-0.5cm}
\end{figure*} 

Figure~\ref{fig:worm_shear}a shows the shear rates at a particular instance during the nematode worm beat cycle for a 2D (left) vs 3D (right) model. Aside from very high shear rates at the worm head and tail, we see that the essential features of the small amplitude solution persist; the 2D sheet generates lower shear near the sheet, takes its maximum ``off'' the sheet, and decays considerably more slowly than the 3D shear rate. This difference is manifest in the effective viscosity arising from such a shear rate. Under the same assumptions and conditions applied to figure~\ref{fig:1}c, the 2D vs 3D fluid viscosity surrounding the finite-length worm is shown in figure~\ref{fig:worm_shear}b. As with the small-amplitude solutions, the relative effective viscosity drops to around $0.2\mu_0$ for the cylinder, although the sheet drops significantly lower. The distance at which the fluid reaches is zero shear rate viscosity, however, remains significantly lower in the 3D vs 2D case, as with figure~\ref{fig:1}c, demonstrating a strong confinement effect in 3D, as noted by G{\'o}mez et al.~\cite{gomez2017helical}, which is not present in the 2D modelling. 

\subsection{Power expenditure scaling law}

{Despite the fact that} the functional forms of the shear rate in 2D and 3D are radically different, Li and Ardekani~\cite{li2015undulatory} recently derived a scaling law based upon the 2D model which has been experimentally verified by Gagnon and Arratia~\cite{gagnon2016cost} for (3D) nematode worms exhibiting planar swimming in shear-thinning fluids. In light of the above, we now briefly explain this agreement. 

The scaling law was derived based upon the assumption used herein, verified by numerical calculation, that the flow shear rate {driven by undulatory swimming} in shear-thinning fluids is approximately equal to that in Newtonian fluids. The experimental validation of Gagnon and Arratia~\cite{gagnon2016cost} indicate that this assumption is also valid for 3D, and so we have the ratio of shear-thinning to Newtonian power ${P_S}/{P_N} ={\int_V \mu|\dot{\gamma}|^2\,\mathrm{d}V}/{\int_V |\dot{\gamma}|^2\,\mathrm{d}V}$, so that
\begin{equation}
\frac{P_S}{P_N} = \frac{\int_V [\mu_s\! +\! (1-\mu_s)(1\! +\!
\mathrm{Cu}^2\dot{\gamma}^2)^{(n-1)/2}]|\dot{\gamma}|^2\,\mathrm{d}V}{\int_V
|\dot{\gamma}|^2\,\mathrm{d}V}, 
\end{equation}
and replacing $\dot{\gamma}$ with a ``typical value'' $\dot{\gamma_T}$~\citep{li2015undulatory} this equation may be written 
\begin{equation}
1 - \frac{P_S}{P_N} = (1-\mu_s) - (1-\mu_s)(1 +
\mathrm{Cu}^2\dot{\gamma_T}^2)^{(n-1)/2}
\quad \Rightarrow\quad \frac{1 - {P_S}/{P_N}}{1-\mu_s} = 1 - (1 +
\mathrm{Cu}^2\dot{\gamma_T}^2)^{(n-1)/2},
\end{equation}
which is the more general form of the scaling law verified experimentally by Gagnon and Arratia~\cite{gagnon2016cost}; in short, this law is only dependent upon the dimensionality of the problem if the typical shear rate $\dot{\gamma_T}$ is estimated from an analytical solution, leading to the $3/8$ constant derived for the 2D case by Li and Ardekani~\cite{li2015undulatory} but not required by Gagnon and Arratia~\cite{gagnon2016cost} due to the availability of experimental data.

\section{Discussion}

{Undulatory propulsion through non-Newtonian fluids is a biomedically-relevant problem that continues to encaptivate many research groups internationally. Strong qualitative agreement between trajectories simulated in 2D and 3D Newtonian fluids suggests that 2D modelling in non-Newtonian fluids should provide a basis for understanding the 3D problem, however dramatic differences in flow derivative fields such as the shear rate show that the translation of 2D non-Newtonian intuition should be handled with care.

This argument was applied specifically to the case of undulatory swimming in shear-thinning fluids.} Experimental~\citep{gomez2017helical} and numerical~\citep{li2015undulatory} evidence suggests that shear-thinning rheology affects undulatory propulsion {via the effective viscosity of the surrounding flow, through a pseudo-confinement effect of a corridor of thinned fluid. In small-amplitude 2D studies, this effect is reversed, with the shear-rate taking its maximum ``off'' the swimmer. At large amplitude in 2D this thinned-corridor can return~\citep{montenegro2013physics,li2015undulatory}, albeit with a likely different form to the 3D case, which may explain why these studies find enhanced swimming speed whereas V{\'e}lez-Cordero and Lauga~\citep{velez2013waving} found no swimming enhancement for an inextensible sheet. Interestingly, the dramatic speed-up predicted by Refs.~\citep{montenegro2013physics,li2015undulatory} and demonstrated experimentally for an artificial bacterial swimmer~\citep{gomez2017helical} is not observed in \textit{C. elegans}~\citep{gagnon2014undulatory}. Two possible reasons for this are 1) difficult to detect, rheology-dependent kinematic changes in the worm, and 2) the fact that in the planar beating case, out-of-plane fluid thinning is likely to be significantly less than in-plane thinning, in effect creating a ``thinned wafer'' of fluid in contrast to a thinned tube~\citep{gomez2017helical}. This provides a strong motivation for the extension of the viscoelastic cylinder framework developed by Fu et al.~\citep{fu2007theory,fu2009swimming} to shear-thinning rheology as an avenue of future study.}

 { This work has focused on shear-thinning rheology, however it is important to note that flow derivative quantities arise when} examining the effects on microscale propulsion of most other non-Newtonian fluids, such as yield stress fluids~\citep{hewitttaylor} via the stress, viscoelastic fluids via polymer stress distribution~\citep{thomases2014mechanisms}, and nematic liquid crystals~\citep{krieger2015microscale}. Whilst 2D modelling remains important, particularly for artificial microswimmers such as Diller's swimming sheet~\citep{diller2014sheet}, addressing how this difference affects our understanding of 3D non-Newtonian swimming for these rheologies is a critical issue for future studies, {providing incentive for the development of new asymptotic theory as well as 3D simulations.}


\begin{acknowledgements} T.D.M-J. is supported by a Royal Commission for the Exhibition of 1851 Research Fellowship. T.D.M-J. would like to thank Paulo Arratia and David Gagnon for experimental data on \textit{C. Elegans} and continued discussions, David Smith for manuscript advice, Eric Lauga for insights and discussion, Thomas Powers for his encouragment with the manuscript, {and the anonymous referees for their helpful feedback.}
\end{acknowledgements}


\end{document}